\documentclass[aps,nofootinbib,superscriptaddress,twocolumn]{revtex4}
\usepackage{graphicx}

\newcommand{\bnmr}{$\beta$-NMR}
\newcommand{\bnqr}{$\beta$-NQR}
\newcommand{\STO}{SrTiO$_3$}
\newcommand{\Li}{$^8$Li}

\begin{document}
\title{Near Surface Phase Transition of \STO\ Studied with Zero Field
$\beta$-Detected Nuclear Spin Relaxation and Resonance}

\author{Z.~Salman}
\affiliation{TRIUMF, 4004 Wesbrook Mall, Vancouver, BC, Canada, V6T 2A3}
\author{R.F.~Kiefl}
\affiliation{TRIUMF, 4004 Wesbrook Mall, Vancouver, BC, Canada, V6T 2A3}
\affiliation{Department of Physics and Astronomy, University of
  British Columbia, Vancouver, BC, Canada V6T 1Z1}
\affiliation{Canadian Institute for Advanced Research, Canada}
\author{K.H.~Chow}
\affiliation{Department of Physics, University of Alberta, Edmonton,
  AB, Canada T6G 2J1}
\author{M.D.~Hossain}
\author{T.A.~Keeler}
\affiliation{Department of Physics and Astronomy, University of
  British Columbia, Vancouver, BC, Canada V6T 1Z1}
\author{S.R.~Kreitzman}
\affiliation{TRIUMF, 4004 Wesbrook Mall, Vancouver, BC, Canada, V6T 2A3}
\author{C.D.P.~Levy}
\author{R.I.~Miller}
\affiliation{TRIUMF, 4004 Wesbrook Mall, Vancouver, BC, Canada, V6T 2A3}
\author{T.J.~Parolin}
\affiliation{Department of Chemistry, University of British Columbia,
  Vancouver, BC, Canada V6T 1Z1}
\author{M.R.~Pearson}
\affiliation{TRIUMF, 4004 Wesbrook Mall, Vancouver, BC, Canada, V6T 2A3}
\author{H.~Saadaoui}
\affiliation{Department of Physics and Astronomy, University of
  British Columbia, Vancouver, BC, Canada V6T 1Z1}
\author{J.D.~Schultz}
\affiliation{Department of Physics and Astronomy, University of
  British Columbia, Vancouver, BC, Canada V6T 1Z1}
\author{M.~Smadella}
\affiliation{TRIUMF, 4004 Wesbrook Mall, Vancouver, BC, Canada, V6T 2A3}
\author{D.~Wang}
\affiliation{Department of Physics and Astronomy, University of
  British Columbia, Vancouver, BC, Canada V6T 1Z1}
\author{W.A.~MacFarlane}
\affiliation{Department of Chemistry, University of British Columbia,
  Vancouver, BC, Canada V6T 1Z1}

\begin{abstract}
  We demonstrate that zero field $\beta$-detected nuclear quadrupole
  resonance (\bnqr) and spin relaxation of low energy \Li\ can be used
  as a sensitive local probe of structural phase transitions near a
  surface. We find that the transition near the surface of \STO\
  single crystal occurs at $T_c \sim 150$~K, i.e.  $\sim 45$~K higher
  than $T_c^{\rm bulk}$, and that the tetragonal domains formed below
  $T_{c}$ are randomly oriented.
\end{abstract}

\maketitle

Strontium Titanate (\STO) is an ionic insulator with remarkable
properties. For example it exhibits ``quantum paraelectricity'' which
can be relieved by oxygen isotope substitution\cite{Dec05F}. It is
also of significant practical importance as a substrate and buffer
layer in electronic heterostructures. At room temperature, bulk
SrTiO$_3$ adopts the cubic perovskite structure, but undergoes an
antiferrodisplacive soft-mode structural phase transition to a
tetragonal phase at $T_c^{\rm bulk} \approx 105$K. Intensive research
on this second order transition has driven many advances in the
general theory of structural phase transitions but has not itself
yielded to complete understanding \cite{Cowley96PTSL}. In \STO, and
also more generally, differences between the phase transitions
occuring in the bulk and those that take place near a free surface are
of considerable scientific and technological interest. For example,
theoretical calculations predict that the structural and magnetic
phase transition temperatures near the surface should be strongly
enhanced \cite{Mills71PRBBinder74PRB}. There are currently relatively
few techniques that can be used to study phase transitions, and in
particular those of a structural nature, near the surface at the
\textsl{local} level. For example, although conventional Nuclear
Magnetic Resonance (NMR) and Nuclear Quadrupole Resonance (NQR) are
widely used to study condensed matter systems, they generally lack the
sensitivity required to investigate surface phenomenon (though notable
exceptions exist \cite{NMRSurface}). However, closely related methods
such as beta detected NMR (\bnmr) or NQR (\bnqr) are considerably more
sensitive \cite{Morris04PRL,Salman04PRB,MacFarlane03PB3,BNMRSurface},
and as a consequence, they are well suited to probe local structural
changes near the surface of interesting materials.

In this paper we present zero magnetic field \Li\ \bnqr\
\cite{Salman04PRB} and spin relaxation measurements near the surface
of a single crystal of \STO. This study demonstrates, for the first
time, that \bnqr\ can be used as a sensitive local probe of structural
phase transitions near a surface. In particular, we find that $T_c
\approx 150$~K near the surface of \STO, considerably higher than
$T_c^{\rm bulk}$.  In addition, we obtain information on the
orientation of the near-surface tetragonal domains below $T_{c}$.
More generally, since many other systems \cite{Salman06PB} produce an
appropriate zero field \bnqr\ signal, we anticipate that our technique
is applicable to a wide class of materials.

As discussed in more detail in Ref.~\cite{Morris04PRL,Salman04PRB},
the \Li\ beam is produced at the isotope separator and accelerator
(ISAC) at TRIUMF in Vancouver, Canada. It is then spin-polarized using
a collinear optical pumping method, yielding nuclear polarization as
high as $70 \%$, and subsequently implanted into the \STO\ sample.
Since the implanted beam energy ($28$ keV) is relatively low, the \Li\
stops at an average depth of $\sim 1500$ \AA. The nuclear
polarization, and its time evolution, is the quantity of interest in
our experiments. It can be measured through the $\beta$-decay
asymmetry, where an electron is emitted preferentially opposite the
direction of the nuclear polarization at the time of decay
\cite{Crane01PRL} and detected by appropriately positioned
scintillation counters. The sample is an $8 \times 10 \times 0.5$~mm
single side epitaxially polished $\langle 100 \rangle$ single crystal
substrate, with RMS surface roughness of $1.5$~\AA\ (Applied
Technology Enterprises). It was mounted on a coldfinger cryostat in an
ultra high vacuum (UHV) environment. Thermal contact between the
sample and coldfinger was achieved with a small amount of UHV
compatible grease (Apiezon N). The temperature gradient between the
sample and diffuser of the cryostat was measured to be less than
$0.2$~K. The \Li\ ions were implanted with their initial polarization
along the surface of the crystal which is perpendicular to the
$\langle 100 \rangle$ direction.

Previous measurements at room temperature \cite{Salman04PRB} have
established that \Li$^+$ occupies three equivalent face center (F)
sites in the cubic \STO\ unit cell, with the Ti$^{4+}$ ions at its
corners. As a consequence of its location, the local symmetry of the F
site is non-cubic even in the cubic phase; therefore, the \Li\
experiences an almost axially symmetric electric field gradient (EFG),
$V_{ij}=\frac{\partial^2 V}{\partial x_i \partial x_j}$, with symmetry
axis normal to the unit cell face\cite{Salman04PRB,MacFarlane03PB3}.
Since the \Li\ nucleus has spin $I=2$ and an electric quadrupole
moment $Q=+33$~mb, it experiences an electric quadrupole interaction.
In zero applied magnetic field, the spin Hamiltonian is given by:
\begin{equation} \label{Hamiltonian}
{\cal H}_q=h\nu_q[{I_z^{2}-2}],
\end{equation}
where $\nu_q=e^{2}qQ/8$ and $eq=V_{zz}$ is the EFG along the symmetry
axis. Thus, the \Li\ nuclear spin polarization, ${\mathbf p}(t)$, can
be highly sensitive to the details of a structural phase transition at
the atomic scale, since one expects such transitions to produce
significant changes in the strength and/or symmetry of the EFG tensor.
In order to better interpret the results of our measurements,
described below, it is useful to keep the following qualitative
behavior in mind: (i) a change in the \textsl{strength} of the EFG
will result in a shift of the resonance frequency, while (ii) a
\textsl{deviation from axial symmetry} of the \Li\ site will introduce
non-axial components and produce an additional term in the
Hamiltonian,
\begin{equation}
{\cal H}_\eta=\eta \nu_q(I_x^2-I_y^2),
\end{equation}
where $\eta=(V_{xx}-V_{yy})/V_{zz}$ is the conventional dimensionless
EFG asymmetry parameter \cite{Cohen57SSP,DasHahn}. Since, in our
experiment, most of the \Li\ are prepared in either the $|+2\rangle$
or $|-2\rangle$ spin states, ${\cal H}_\eta$ introduces mixing between
the $|\pm 2 \rangle$ spin states with a characteristic frequency
splitting $\Delta_{\pm 2} \simeq 3 \eta^2 \nu_q$ \cite{Salman04PRB}.
This results in a loss of polarization on the timescale of
$1/\Delta_{\pm 2}$.

\begin{figure}[h]
\includegraphics[width=\columnwidth]{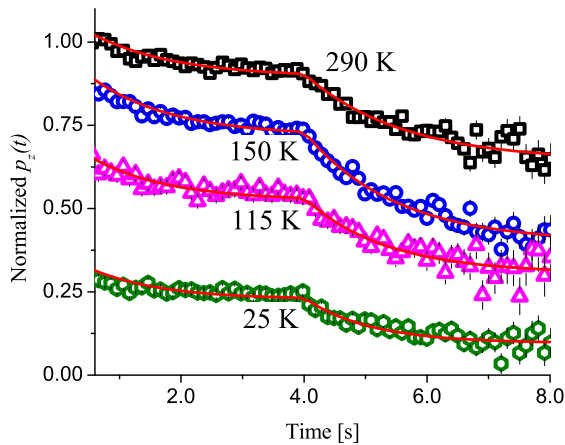}
\caption{The polarization as a function of time, normalized to its
  initial value at $T=290$~K, at several temperatures.}
\label{Relaxation}
\end{figure}

\begin{figure}[h]
\includegraphics[width=\columnwidth]{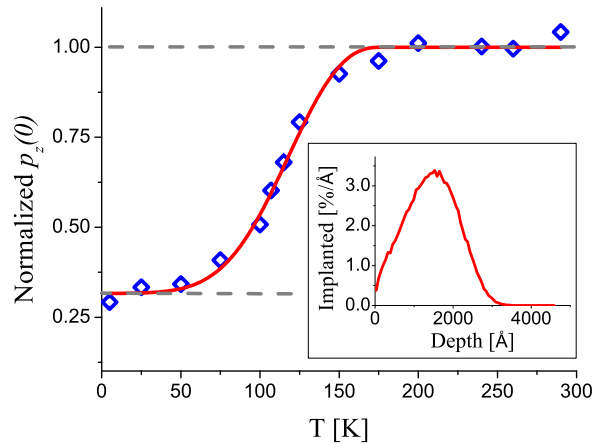}
\caption{The normalized initial polarization as a function of
  temperature. The solid line is a guide to the eye. The inset is the
  calculated \Li\ stopping profile.}
\label{Polarization}
\end{figure}

First, we discuss the observed temperature dependence of ${\mathbf
  p}(t)$ in the \STO\ sample. The polarization along $\hat{z}$,
$p_z(t)$, is measured using a method where a pulse of \Li\ is
implanted at a rate of about 10$^6$/s, starting at $t=0$ for a period
$T=4$ seconds, and the $\beta$-decay asymmetry both during and after
the beam period is measured. Measurements of $p_z(t)$ at different
temperatures are shown in Fig.~\ref{Relaxation}.
The initial polarization was normalized to its value at room
temperature. $p_z(t)$ is determined by both the \Li\ spin-lattice
relaxation rate $\lambda$ and its radioactive lifetime $\tau=1.21$s.
Assuming a general spin relaxation function $f(t,t_p:\lambda)$ for the
fraction of \Li\ implanted in the sample at $t_p$, the polarization
follows
\begin{equation} \label{genrlx}
p_z(t)=\left\{
\begin{array}{ll}
\frac{\int_0^t e^{-(t-t_p)/\tau} f(t,t_p:\lambda)
  dt_p}{\int_0^t e^{-t/\tau} dt} & t \le T \\
\frac{\int_0^T e^{-(T-t_p)/\tau} f(t,t_p:\lambda)
  dt_p}{\int_0^T e^{-t/\tau} dt} & t > T .
\end{array}
\right.
\end{equation}
The data in Fig.~\ref{Relaxation} are fit to Eq.~\ref{genrlx} with a
phenomenological bi-exponential form,
\begin{equation}
f(t,t_p:\lambda)=A_1e^{-\lambda_1(t-t_p)}+A_2e^{-\lambda_2(t-t_p)}.
\end{equation}
The relaxation rates from the fits are small and do not vary much with
temperature. In contrast, as shown in Fig.~\ref{Polarization}, the
initial polarization $p_z(0)$ exhibits a strong temperature
dependence. At high temperatures, $p_z(0)$ is constant and close to
unity. However, it decreases dramatically below $150$~K, eventually
reaching a temperature independent value of $\sim 1/3$ below $75$~K.
As discussed earlier, we attribute the loss of initial polarization to
the appearance of non-axial components of the EFG which arise as the
crystal's symmetry is lowered below the phase transition. The most
striking result in this work is that the loss of polarization in
Fig.~\ref{Polarization} starts at $T_c \simeq 150$~K, much higher than
$T_c^{\rm bulk}$. We attribute this enhancement of $T_c$ to proximity
to the free surface of the crystal.

\begin{figure}[h]
\includegraphics[width=\columnwidth]{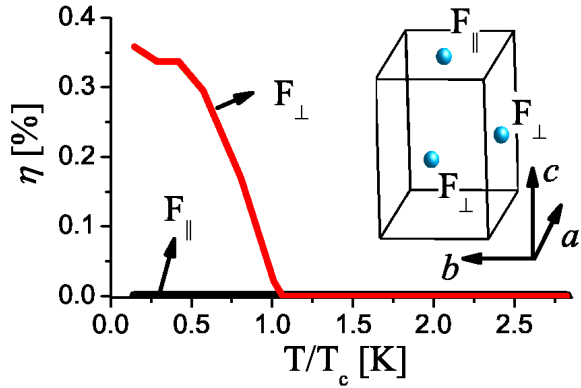}
\caption{The calculated $\eta$ as a function of temperature. The inset
shows the different types of \Li\ sites in \STO\ unit cell.}
  \label{Eta}
\end{figure}

In order to fully understand the results in Fig.~\ref{Relaxation} and
Fig.~\ref{Polarization} one should consider the changes in the local
structure around the \Li\ below the transition, where the lattice
parameter $c$ becomes larger than $a$ (see inset of Fig.~\ref{Eta}).
Therefore, two inequivalent \Li\ sites now exist: those where the
$c$-axis is perpendicular to the EFG's principal axis (F$_{\perp}$)
and those (F$_{\parallel}$) where it is parallel. These sites occur in
a 2:1 ratio (F$_{\perp}$:F$_{\parallel}$) and hence 2/3 of the \Li\
sites experience non-axial distortions in the EFG. Treating the
lattice of \STO\ as an array of point charges (Sr$^{2+}$, O$^{2-}$ and
Ti$^{4+}$), we calculated $\eta$ for both types of F site as a
function of temperature, using the bulk lattice constants reported in
Ref.~\cite{Okazaki73MRB}. As can be seen in Fig.~\ref{Eta}, $\eta$ is
exactly zero for both types in the cubic phase. It remains unchanged
through the phase transition for the F$_{\parallel}$ site, while for
the F$_{\perp}$ sites, it increases up to $\sim 0.35\%$ due to the
tetragonal distortion. The increase in $\eta$ at $T_{c}$ causes an
increase in $\Delta_{\pm 2}$ which is sufficient to produce a partial
loss of $p_{z}(0)$ due to \Li\ in F$_{\perp}$ sites.

\begin{figure}[h]
\includegraphics[width=\columnwidth]{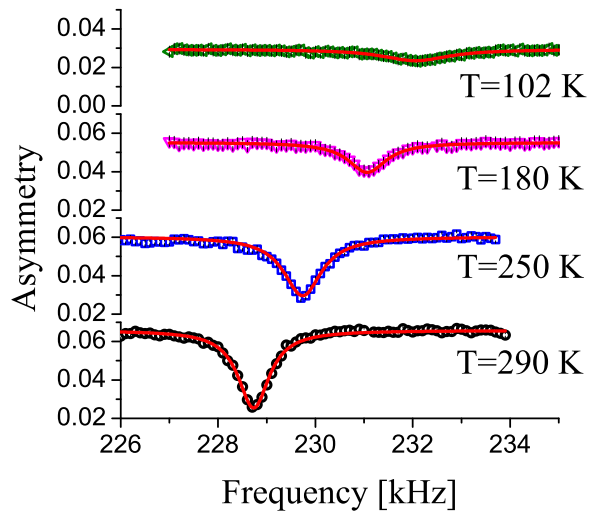}
\caption{\Li\ \bnqr\ lines in \STO\ for different temperatures. The
  line shifts, broadens and weakens as the sample is cooled from room
  temperature to $100$ K. The solid lines are fits to a Lorentzian,
  which describes the data very accurately.}
\label{Lines}
\end{figure}

\begin{figure}
\includegraphics[width=\columnwidth]{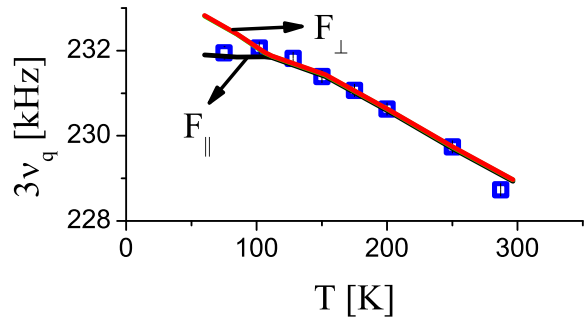}
\caption{The resonance frequency, $3 \nu_q$, as a function of
  temperature, obtained from fitting the lines to a Lorentzian shape.
  The solid lines are the calculated value of $3 \nu_q$ from the
  relative increase of $\nu_q$ scaled to match the data at $250$~K.}
  \label{Freq}
\end{figure}

Additional information, needed to estimate $\Delta_{\pm 2}$, can be
obtained by measuring the value of $3\nu_q$ using the nuclear
quadrupole \textsl{resonance}, as detailed in
Ref.\cite{Salman04PRB}. This was carried from room temperature down to
$\sim 75$~K, below which the resonance could not be observed.
Representative spectra at several temperatures are shown in
Fig.~\ref{Lines}. As can be seen in Fig.~\ref{Freq}, the resonance
frequency, which occurs at $3 \nu_q$, increases down to $\sim 100$~K,
below which it appears to saturate. This temperature dependence
reflects an increase in the EFG at the \Li\ site, which is consistent
with thermal contraction of the \STO\ lattice, as confirmed by the
comparison with simulations using the point charge model.
While the absolute value of $3 \nu_q$ obtained from such a simple
calculation is generally not expected to agree well with measurements
\cite{Cohen57SSP}, the relative increase in $3\nu_q$ as a function of
temperature (solid lines in Fig.~\ref{Freq}) yields very good
agreement with the experiment from $290$~K down to $105$~K. The
bifurcation of the calculated values at $T=105$~K is due to the bulk
phase transition; while the calculated value of 3$\nu_q$ for the
F$_{\parallel}$ site becomes temperature independent, it continues to
increase for the F$_{\perp}$ sites. The experimental results agree
with the calculations for the F$_{\parallel}$ site only, and there is
no evidence for a signal from the F$_{\perp}$ sites. This is further
confirmation that below the transition, $\eta$ for the F$_{\perp}$
sites becomes significant (see Fig.~\ref{Eta}); consequently, the
polarization of \Li\ in these sites is lost and does not contribute to
the resonance. Note that a typical value of $3 \nu_q \simeq 230$~kHz
provides a sufficiently large $\Delta_{\pm 2}$ for $\eta$ as low as
$0.1 \%$ to produce the observed loss of polarization.

It is important to point out that since $2/3$ of the polarization of
\Li\ stopping in tetragonal domains (where the phase transition has
occurred) is lost, the main contribution to the resonance comes from
\Li\ stopping in the cubic phase. Therefore the resonances are less
sensitive to the structural phase transition near the surface, since
at $T \gtrsim T_c^{\rm bulk}$ they will be dominated by \Li\ stopping
far from the surface. This is evident from the fact that the
calculated resonance frequency, using {\em bulk} lattice constants,
yields good agreement with the experiment (see Fig.~\ref{Freq}), and
that the resonance amplitude becomes very small below $T_c^{\rm
  bulk}$.

As mentioned earlier, our observed $T_{c} \simeq 150$~K near the
surface of \STO\ is significantly higher than $T_{c}^{\rm bulk}$,
corresponding to a difference of $\Delta T_{c} \approx 45$K. Based on
an extrapolation of the penetration depth dependence of the x-ray
scattering parameters, an increase of $\Delta T_c=220$~K at the
surface was predicted \cite{Osterman88JPC}. Optical second harmonic
generation (SHG) studies recently found that the phase transition near
the surface of \STO\ occurs $\Delta T_c=45$~K above the temperature of
the bulk phase transition \cite{Mishina00PRL}. A similar result was
found closer to the surface by electron diffraction
\cite{Krainyokova96CJP}. Unlike these techniques, the \Li\ nuclei
sense the phase transition at the atomic scale, while the net signal
is averaged over \Li\ sites in the implantation volume, which is a
beamspot about $3$ mm in diameter together with an implantation depth
profile. As shown in the inset of Fig.~\ref{Polarization}, Monte
Carlo calculations using the TRIM.SP \cite{TRIM} package predicts that
\Li\ has an average implantation depth of $\sim 1500$~\AA, a width
(range straggling) of $\sim 2000$~\AA, and a maximum depth of $\sim
3000$~\AA. Thus the measured $p_z(t)$ spectra are composed of signals
from \Li\ stopping at varying distances from the surface. The lower
symmetry at the surface (together with effects such as surface
reconstruction \cite{STOSurface}) certainly presents a perturbative
influence on the bulk structural phase transition. A simple model of
this effect is to assume that $T_c$ is a monotonically decreasing
function of depth, falling from a maximal value $T_c^{\rm surf}$ at
the surface to $T_c^{\rm bulk}$ at large depths. In this picture, the
observed breadth of the transition is due, at least in part, to a
weighted averaging over the intrinsically inhomogeneous $T_c$
distribution. The remarkable similarity of our estimate of $T_c^{\rm
  surf}$ to that of the surface-related reflection SHG signal
\cite{Mishina00PRL} suggests a common intrinsic origin to the
enhancement of $T_c$.

Finally, since $p_{z}(t)$ is strongly dependent on the direction of
the $c$-axis when the sample undergoes a transition into the
tetragonal phase, it provides information on the orientation of the
tetragonal domains near the surface. In particular, the observed
$p_{z}(0)$ is non-zero at low temperatures in the tetragonal
phase. This indicates that a fraction of the implanted \Li\ still
experiences an axially symmetric ($\eta \simeq 0$) EFG along its
polarization, corresponding to \Li\ in the F$_{\parallel}$ site. The
observed $1/3$ value is strong evidence that the $c$-axis of the
tetragonal domains is oriented randomly, in agreement with
Ref.~\cite{Buckley99JAP}.

In conclusion, we have demonstrated that \bnqr\ can be used to study
structural phase transitions near the surface of \STO. The transition
occurs at $\sim 150$K, compared to $105$~K in the bulk. The tetragonal
domains that are formed at low temperatures were found to be randomly
oriented. Analogous studies to those reported here, but at different
implantation energies (and therefore stopping depths) will allow
depth-profiling of this surface proximity effect. We are currently
augmenting our \bnqr\ spectrometer with a high voltage platform
\cite{Morris04PRL} in order to enable such a depth-resolved study. We
emphasize that the techniques described in this paper are not
restricted to the study of \STO, but should be applicable to
investigations of structural phase transitions in many other materials
where a nuclear quadrupole interaction is observed \cite{Salman06PB}.

This work was supported by the CIAR, NSERC and TRIUMF. We thank Rahim
Abasalti, Bassam Hitti and Donald Arseneau for technical support. We
also thank Laura Greene for providing the SrTiO$_3$ crystal, W.
Eckstein for providing the TRIM.SP code, and K.-C. Chou and W.J.L.
Buyers for helpful discussions.

\newcommand{\noopsort}[1]{} \newcommand{\printfirst}[2]{#1}
  \newcommand{\singleletter}[1]{#1} \newcommand{\switchargs}[2]{#2#1}

\end{document}